# Rejoinder: Struggles with Survey Weighting and Regression Modeling

**Andrew Gelman**

## 1. MOTIVATIONS

I was motivated to write this paper, with its controversial opening line, "Survey weighting is a mess," from various experiences as an applied statistician:

- Encountering this sort of statement in the documentation of opinion poll data we were analyzing in political science: "A weight is assigned to each sample record, and MUST be used for all tabulations." (This particular version was in the codebook for the 1988 CBS News/New York Times Poll; as you can see, this is a problem that has been bugging me for a long time.) Computing weighted averages is fine, but weighted regression is a little more tricky—I do not really know what a weighted logistic regression likelihood, for example, is supposed to represent.
- Constructing the weighting for the New York City Social Indicators Survey (SIS). It quickly became clear that we had to make many arbitrary choices about inclusion and smoothing of weighting variables, and we could not find any good general guidelines.
- We wanted to estimate state-level public opinion from national polls. If our surveys were simple random samples, this would be basic Bayes hierarchical modeling (with 50 groups, possibly linked using state-level predictors). Actually, though, the surveys suffer differential nonresponse (lower response by men, younger people, ethnic minorities, etc.) as signaled to the user (such as myself) via a vector of weights.

*Andrew Gelman is Professor of Statistics and Professor of Political Science, Department of Statistics, Columbia University, New York, New York 10027, USA e-mail: gelman@stat.columbia.edu.*

The weighting in others' surveys, as well as our own SIS, appeared to be a mess. In particular, different survey organizations weight on different variables, and use different smoothing of weights, even when using similar methodology to survey the same population (Voss, Gelman and King, 1995). The weights are clearly not the platonic inverse-selection probabilities envisioned in some of the classical statistical theory of sampling.

Having established that survey weighting is a mess, I should also acknowledge that, by this standard, regression modeling is also a mess, involving many arbitrary choices of variable selection, transformations and modeling of interaction. Nonetheless, regression modeling is a mess with which I am comfortable (Gelman and Hill, 2007) and, perhaps more relevant to the discussion, can be extended using multilevel models to get inference for small cross-classifications or small areas.

I was thus motivated to get the benefits of weighting—adjusting for expected or known differences between sample and population—in the familiar and expandable context of regression modeling. As indicated by the title of the paper, we are not there yet. I am thrilled to have my paper discussed by leaders in survey research who have made so many important contributions in the theory and practice of survey analysis, and I hope this discussion helps us move the field forward, both toward my ultimate goal of a unified design-based and model-based analysis, and toward the intermediate goal of identifying weak points of currently used weighting and poststratification adjustments.

## 2. SAMPLE SELECTION PROBABILITIES

Unequal probabilities of inclusion in a survey arise in three ways: stratification or multiple frames (so that units in different strata have different selection probabilities, perhaps unavoidably or perhaps by design), clustering and nonresponse. Unfortunately, surveys sometimes simply supply a weight without explaining where it came from. This can allow consistent estimates using weighted regression, but, as

TABLE 1
*Proportion of U.S. households and adult population in households of different sizes, along with weighting adjustments for surveys that sample households at random, followed by sampling one adult per selected household*

| Number of adults in household | Proportion of households (census) | Proportion of adults in each type of household (census) | Weights from theory | Weights from poststratification | | |
|---|---|---|---|---|---|---|
| | | | | CBS.1 | CBS.2 | NES |
| 1 | 0.35 | 0.20 | 1 | 1 | 1 | 1 |
| 2 | 0.55 | 0.62 | 2 | 1.32 | 1.38 | 2.00 |
| 3 | 0.08 | 0.13 | 3 | 1.35 | 1.53 | 2.30 |
| 4+ | 0.02 | 0.05 | 4.25 | 0.95 | 1.20 | 2.55 |

From Gelman and Little (1998). Theoretical inverse-probability weights are proportional to household size. Poststratification weights—shown for three different national pre-election polls from 1988—more accurately match the sample to the population and capture the increased probability of response in larger households. The weights from the National Election Survey ("NES") are closer to the theoretical weights, which makes sense since that survey puts in extra effort to reach people and has a lower nonresponse rate. For the purpose of this article, the point of this table is to illustrate that, in surveys of human populations, poststratification adjustments can sometimes be more important than selection probabilities.

several discussants note, design knowledge is needed in order to correctly compute standard errors.

Nonresponse can contaminate selection probabilities that otherwise would be simple. For example, Table 1 shows nominal inverse-probability weights and actual poststratification weights for households of different sizes from two different national pre-election surveys from 1988, each of which was performed by selecting households at random and then, in each sampled household, selecting an adult at random. The nominal inverse-probability weight for a household is simply the number of adults in the household (with a slight adjustment at the top cell, where we are combining all households with four or more adults). The poststratification weights are actually less variable—it turns out that it is easier to reach someone in a household with many adults, and this nonresponse partially cancels the probability correction. (The actual surveys used either the nominal weights or no weighting at all, leading to weighted samples that were highly unrepresentative of the population of household sizes. As discussed in Gelman and Little, 1998, this error had minimal impact because household size has a near-zero correlation with political preference—households with two adults are more Republican, and those with one or three adults tend to support the Democrats.)

In the household size example, there is no advantage to doing the inverse-probability weights since the poststratification weights overwrite them anyway. (In a survey where it would be too much trouble to get the Census numbers to poststratify, Gelman and Little, 1998, recommend using the square root of household size as an approximate weighting factor.)

It is common to confuse poststratification weights with inverse-selection probabilities; for example, Pfeffermann in his otherwise excellent discussion writes, "An alternative procedure [to full regression modeling] favored by survey analysts is to... use weighted regression with the weights defined by the inverse of the sample inclusion probabilities." But in all the surveys of human populations with which I have worked, key parts of the weights are defined by poststratification. This is not to dismiss Pfeffermann's comments—in particular, weighted regression gives consistent estimates if weights are defined by poststratification. It is just worth pointing out that statistical theory often falls into the trap of assuming a sampling design (random sampling with known unequal selection probabilities) that is not always realistic.

## 3. TWO GOOD EXISTING APPROACHES TO ANALYSIS

As several discussants note, my paper was too pessimistic by not acknowledging two reasonable and currently available methods for analyzing survey data in the presence of design complications (mostly stratification and clustering) and nonresponse: the first approach uses unit weights, the second uses modeling and poststratification. Both methods have loose ends but can work reasonably well in practice (which is why I characterized survey weighting as a "mess," not a "disaster").

The first approach begins with the construction of weights. Some of these weights can be created at the design stage, others after the data have been collected. In my own experience, the design weights have been the most important in surveys of records (e.g., samples of insurance claims, where the largest claims are, by design, disproportionately likely to be picked), and poststratification weights have been the most important in surveys of people (where the typical goal is to represent a population where all people are counted equally).

Weighting typically requires some choices, but, as noted in the discussions by Pfeffermann and by Breidt and Opsomer, survey researchers in practice have been able to come up with numerically stable weights that correct for important differences between sample and population. Weighting may be more an art than a science, but it is an art with many successful practitioners. As Bell and Cohen explain, criteria such as mean squared error can be used to evaluate different choices of weighting.

When computing summaries more complicated than means and quantiles, weights can be incorporated into inference as factors in the log-likelihood for each observation, which reduces, for example, to weighted linear or logistic regression (Pfeffermann, 1993). For maximizing the likelihood, only relative weights are necessary. Standard errors are more of a challenge, but as several discussants pointed out, they can be computed using methods such as the jackknife or bootstrap which subdivide or resample the data according to the stratification, clustering and poststratification in the design and estimate. Binder (1983) is an important paper in this area, and Lu and Gelman (2003) reports our own adaptation of this approach.

The second generally successful approach of survey analysis uses regression modeling that includes, as predictors, all the variables that affect the probability of inclusion (and thus would be used in constructing the survey weights). Having fit such a model, population-average inferences can be averaged over using the joint population distribution of these predictors, from either a census or some estimation procedure such as iterative proportional fitting (Deming and Stephan, 1940). As Little notes in his discussion, nonparametric methods such as the bootstrap can also be used to construct, or to check, variance estimators for these models.

As discussed in my paper, the modeling approach can be challenging because of the large number of potential interactions; however, this method has had much success, especially for small-area estimation.

## 4. DESIGN-BASED AND MODEL-BASED

I avoid the usual characterization of survey weights as "design-based" or "model-based" because, as Little points out, both the above approaches are fundamentally design-based in that they make use of the information used in the design. I do not think a purely model-based approach is appropriate except for very simple surveys or as a starting point until design information can be incorporated into the model.

For example, in a recent project on estimating state-level opinions from a series of national polls (Zhou and Gelman, 2007), we started with a simple small-area estimation model just using the number of Yes and No responses in each state in each year, putting some effort into fitting a hierarchical model to the 50 state-level time series. Once we had succeeded in setting up a reasonable model and fitting it—not a trivial task for us, but having nothing to do with sample-survey issues—we took the next step and modeled the responses conditional on demographic variables (sex, ethnicity, age, etc.) as well as state and year. This larger model takes account of the design—in this case, adjusting for variables that are predictive of nonresponse—but takes more work, first in the modeling stage and then because the inferences must be added over cells by poststratifying on Census totals for each year. The preliminary, purely model-based analysis is extremely helpful to us in building our model, but for substantive conclusions we need the adjusted and poststratified analysis which accounts for possible systematic differences between sample and population.

Ultimately we need ways of evaluating methods that stand apart from the design-based/model-based distinction. I like Lohr's general set of principles by which we can evaluate different methods.

## 5. POTENTIAL BENEFIT FROM A UNIFIED APPROACH

Given the existence of two serviceable methods, why am I still struggling with survey weighting and regression modeling?

One reason for my struggle is that the unit weighting (which is the state-of-the-art method for population inference) is not completely compatible with hierarchical modeling and poststratification (the state-of-the-art method for small-area estimation). This falls in Lohr's criteria of internal consistency and calibration.

Another reason I am struggling is that I am trying to solve difficult small-area estimation problems—for example, state-level time trends in support for the death penalty—which put more burden on our modeling assumptions. To go in the other direction, I am estimating population averages—for example, the change in average satisfaction about the public schools among New Yorkers with children—and these rely on weighting procedures that were not really constructed for the purpose of estimating changes.

Perhaps because of my general experience with hierarchical modeling, I am more comfortable with a model-based approach—for all its difficulties with interactions—but I would be happy to use weights if I felt I could really trust the results. One potential benefit of a unified approach is that, if we have a hierarchical model we think is reasonable, we can back out its equivalent weights (as in Section 3.2 of my paper) and then evaluate and use these weights as would be done using classical weighting methods. If it is unacceptable to have different weights for different outcome measures $y$, we could average the sets of weights estimated for different outcomes of interest.

As noted by Lohr, the resulting weights can be negative—in this case, I would think there is a problem with the model. If the model could not be fixed, it might be appropriate to set the negative weights to zero. As Little notes, there is a duality between collapsing or smoothing weights and removing or smoothing interactions in a regression model. I would assume that, in many cases, negative weights would correspond to nonsensical regression models. On the other hand, if the true relation between $y$ and $x$ has nonlinearities, negative weights could make sense (e.g., if a particular subpopulation were known or estimated to be the opposite of the majority).

It would also be desirable to use nonlinear methods such as regression trees (as mentioned by Bell and Cohen), but then it would seem difficult to construct even approximately equivalent weights. Weighting and fully nonlinear models would seem to be completely incompatible methods.

## 6. CHALLENGES WITH REGRESSION OR WEIGHTING

Pfeffermann raises several areas where modeling (or weighting) can get difficult. First, what happens with the regression approach when there are empty cells? In this case, it is necessary to exclude some interactions or shrink them using a multilevel model, and indeed I would do so. My article discusses simple regression as a reference point because, for the fully saturated model, it reduces to simple weighting. (In this case, the cell sizes $N$ are indeed additional information, as Pfeffermann notes.)

Pfeffermann points out that cell sizes are not always known. We agree that, in this case, variance estimation for the model-based approach becomes more complicated. Ultimately we would like a fully Bayesian approach which averages over the posterior distribution of the cell sizes—yet another step in our "struggle."

Pfeffermann also emphasizes that our method assumes missingness at random. We are definitely working within the classical tradition in which a survey is adjusted based on some fixed set of fully observed variables, and nonignorable missingness is a problem with both of the current standard methods described in Section 3 above. In theory, the model-based approach could be expanded to include a nonignorable model, but we have not actually done this in our examples. As has been emphasized by Little, the inclusion of more variables in the weighting (or poststratification) potentially brings the missing-at-random model closer to realism. At this point, additional effort is required to model the survey response given all these predictors (in our approach) or to deal with weights that depend on many predictors (in Pfeffermann's approach). We would hope that our equivalent weights would be close to those obtained by Pfeffermann, at least in linear or approximately linear response models in which estimates based on weighted averages of the data make sense.

## 7. SUMMARY

Ultimately, I see weighting as a powerful tool but with serious limitations, especially for small-area estimation because weighting cannot handle empty cells. I like the idea of equivalent weights as a unification with regression modeling (possibly also, as Little suggests, linking to the inverse-propensity-weighting approach of Scharfstein, Rotnitzky and Robins, 1999). However, equivalent weights have their own difficulties, especially for nonlinear models. As Breidt and Opsomer emphasize, the design-based paradigm is crucial—one question is how best to adapt it to challenging problems such as estimation of small subpopulations and time series comparisons where underlying populations are changing.

## ACKNOWLEDGMENTS

We thank Ed George for organizing the discussion and the National Science Foundation and National Institutes of Health for financial support.

## REFERENCES

- BINDER, D. A. (1983). On the variances of asymptotically normal estimators from complex surveys. *Internat. Statist. Rev.* **51** 279–292. MR0731144
- DEMING, W. E. and STEPHAN, F. F. (1940). On a least squares adjustment of a sampled frequency table when the expected marginal totals are known. *Ann. Math. Statist.* **11** 427–444. MR0003527
- GELMAN, A. and HILL, J. (2007). *Data Analysis Using Regression and Multilevel/Hierarchical Models*. Cambridge Univ. Press.
- GELMAN, A. and LITTLE, T. C. (1998). Improving on probability weighting for household size. *Public Opinion Quarterly* **62** 398–404.
- LU, H. and GELMAN, A. (2003). A method for estimating design-based sampling variances for surveys with weighting, poststratification and raking. *J. Official Statistics* **19** 133–151.
- PFEFFERMANN, D. (1993). The role of sampling weights when modeling survey data. *Internat. Statist. Rev.* **61** 317–337.
- SCHARFSTEIN, D. O., ROTNITZKY, A. and ROBINS, J. M. (1999). Adjusting for nonignorable drop-out using semiparametric nonresponse models (with discussion). *J. Amer. Statist. Assoc.* **94** 1096–1146. MR1731478
- VOSS, D. S., GELMAN, A. and KING, G. (1995). Preelection survey methodology: Details from eight polling organizations, 1988 and 1992. *Public Opinion Quarterly* **59** 98–132.
- ZHOU, S. and GELMAN, A. (2007). Estimating time series of state-level opinions from national polls. Technical report, Dept. Statistics, Columbia Univ.